\documentclass[a4paper,11pt]{article}
\usepackage{jinstpub} % for details on the use of the package, please see the JINST-author-manual
\usepackage{lineno}
\usepackage{color}
\usepackage{mathtools}
\usepackage[normalem]{ulem} % cancel line
\title{\boldmath Radon Concentration Measurement with a High-Sensitivity Radon Detector at the Yemilab}

\author[a]{Kyung Min Seo,}
\author[a]{Hyunsoo Kim,}
\author[b,c]{Yeong Duk Kim,}
\author[b]{Hye Young Lee,}
\author[b]{Jaison Lee,}
\author[b,c,1]{Moo Hyun Lee,\note{Corresponding author.}}
\author[b]{Jungho So,}
\author[b]{Sang Cheol Yoon,}
\author[d]{and Young Soo Yoon}
\affiliation[a]{Department of Physics and Astronomy, Sejong University, Seoul, Republic of Korea}
\affiliation[b]{Center for Underground Physics, Institute for Basic Science (IBS), Daejeon, Republic of Korea}
\affiliation[c]{IBS School, University of Science and Technology (UST), Daejeon, Republic of Korea}
\affiliation[d]{Korea Research Institute of Standards and Science (KRISS), Daejeon, Republic of Korea}
\emailAdd{mhlee@ibs.re.kr}

\abstract{
The radiation emitted from radon is a critical background in rare event search experiments conducted at the Yemi Underground Laboratory (Yemilab) in Jeongseon, Korea.
A Radon Reduction System (RRS) has been developed and installed in Yemilab to reduce radon concentration in the air.
The RRS primarily provides a purified air of $\rm 50\ m^{3}/h$ to the cleanroom used to assemble crystal detectors in the AMoRE, a neutrinoless double beta decay search experiment.
RRS can reduce the radon level by a factor of 300, so a high-sensitivity radon detector was required.
A highly sensitive radon detector was constructed using a 70 L chamber with a large PIN photodiode to measure radon concentration in the purified air. 
The radon detector shows an excellent resolution of 72 keV (FWHM) for 6.003 MeV alphas from {$^{218}$}Po decay and a sensitivity down to 23.8 $\pm$~2.1 mBq/m{$^3$} with a boil-off N{$_2$} gas sample.  
The radon concentration level from the RRS measured by the radon detector was below ${\rm 0.29 ~Bq/m{^3}}$ with a reduction factor of about 300.
}

\begin{document}
\maketitle
\flushbottom

\section{Introduction}
\label{sec:intro}

%The current standard model, completed by the Higgs particle discovered in 2012, has improved our understanding of the universe by defining the fundamentals of matter. However, there are experimental facts that cannot be explained by these standard models: the theory of neutrino mass, the matter-antimatter asymmetry of the universe, etc. These problems suggest a new physics that goes beyond the current standard model. Since these experimental facts have very rare observational probabilities, understanding and reducing the experiment's background is the most critical task to achieve the purpose of the experiment. On the ground, the signal we want is diluted by the cosmic rays like a muon from the cosmos, making it challenging to observe. However, if the experimental facility is installed underground, the ground surface becomes a shield from the cosmic rays, and the background can be dramatically reduced. A laboratory with an overburden of about 700 m in Korea has been built in Yangyang, Gangwon-do, and a new underground laboratory with an overburden of 1,400 m is being constructed in Jeongseon.

\begin{figure}
    \centering
    \includegraphics[width=0.9\textwidth]{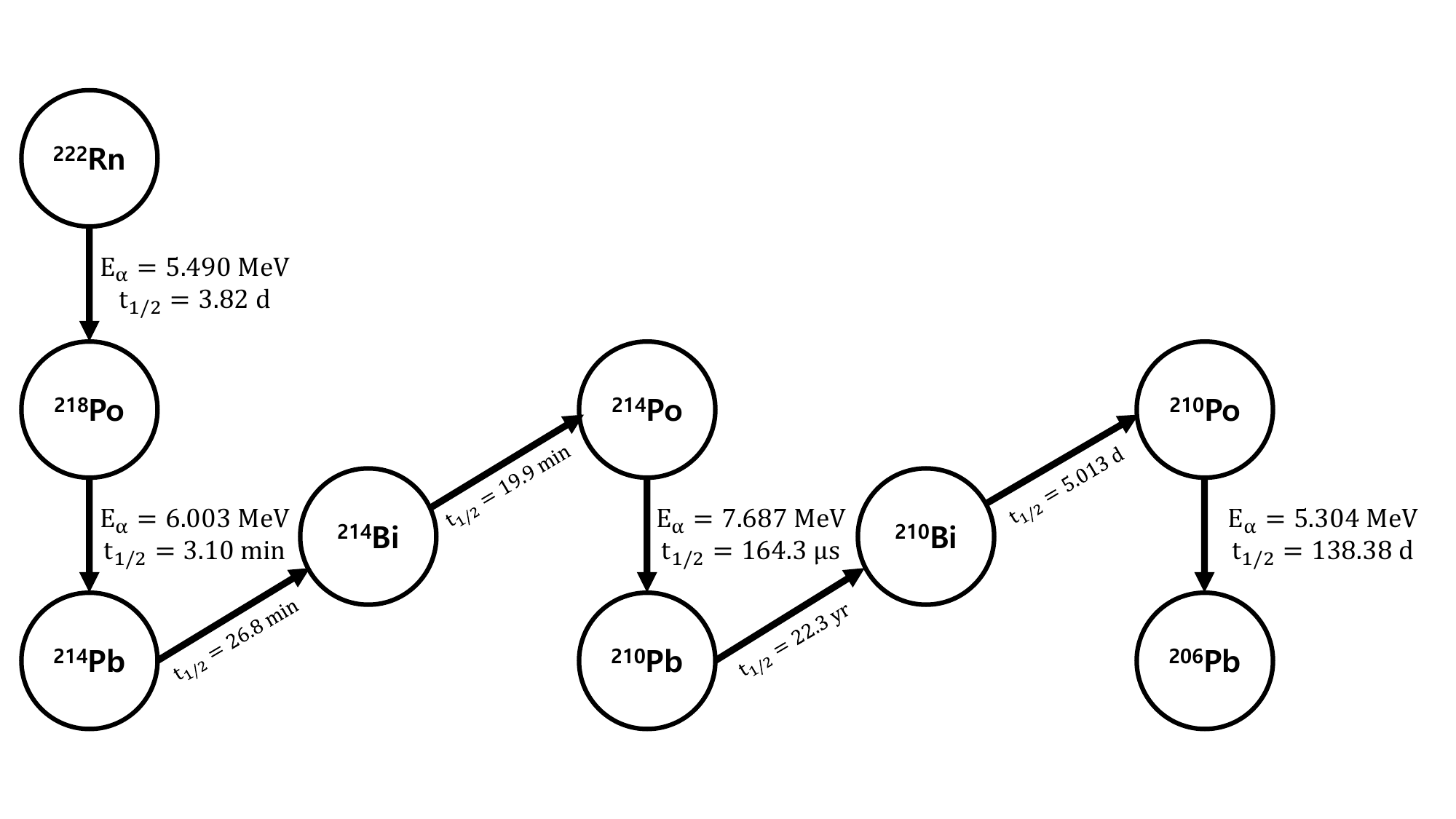}
    \includegraphics[width=0.9\textwidth]{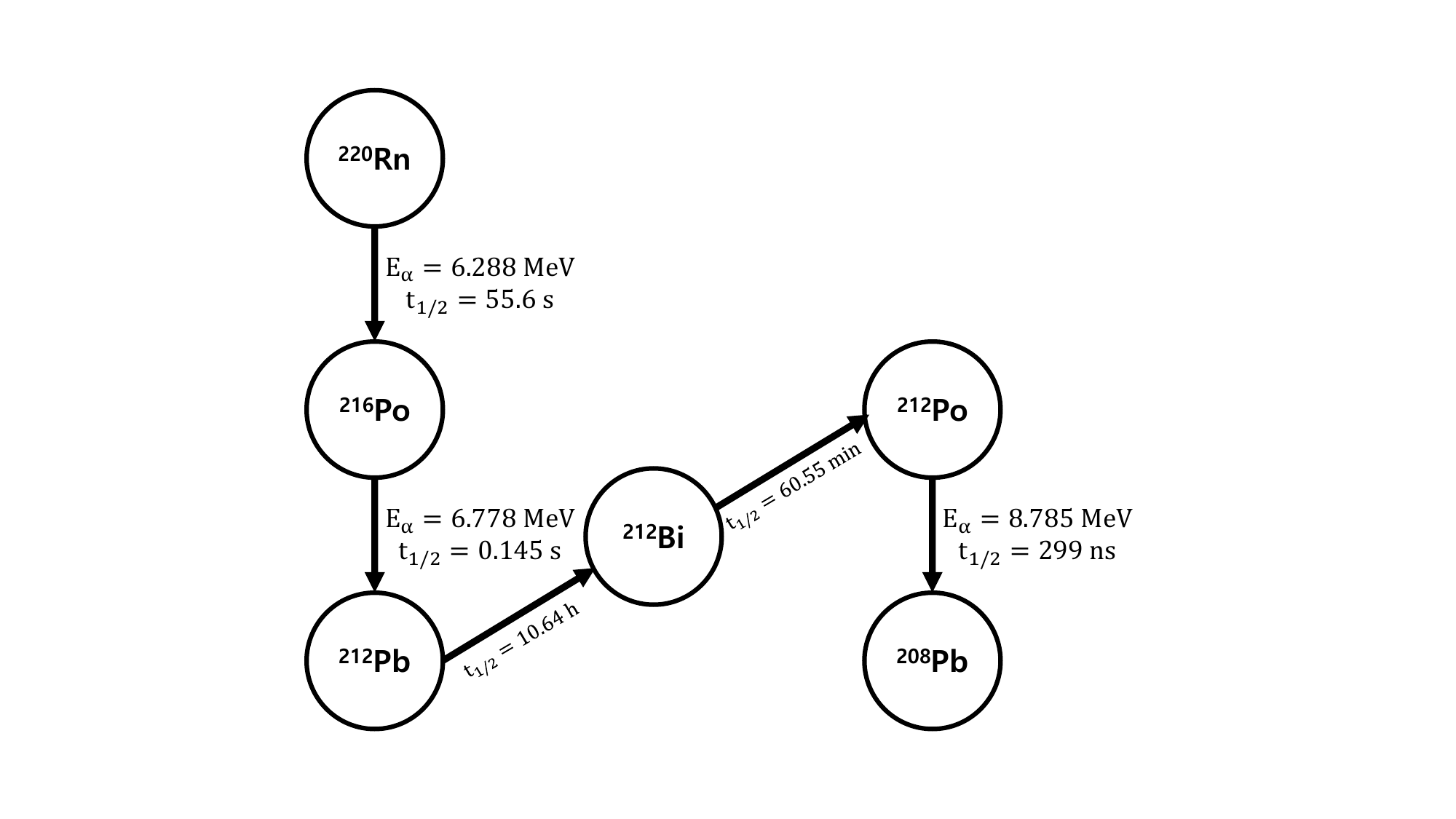} 
    \caption{A schematic of Radium series from the $\rm ^{238}U (top)$ and $\rm ^{232}Th (bottom)$ decay chain. For $\beta$ decay, only the half-life information was included.}
    \label{fig:RSeries}
\end{figure}

%Among the elements constituting the Earth's surface, uranium and thorium have radium as a daughter nucleus. Radium undergoes radioactive decay, producing radon.
The underground laboratory, developed to reduce the background caused by cosmic rays, must consider the other background created by the Earth's surface, radon. Radon is derived from uranium and thorium, commonly present in the Earth's crust, and can be represented by the radium series from $\rm ^{238}U$ and $\rm ^{232}Th$ decay chains, as shown in Fig. \ref{fig:RSeries}.
Radon is a radioactive noble gas with very low chemical reactivity in general conditions. 
%\textcolor{red}{The reason for low chemical reactivity is that the outermost electron shell of the atom is full. (?? jslee)}
Because of these characteristics, the concentration of radon cannot be measured directly.
However, the radon concentration can be inferred through the daughter nuclide, poloniums.
%Polonium is a radioactive metal with a positive charge, which allows it to be captured from the air via an electric field.
Subsequent stages of radon decay ultimately result in lead production, accompanied by the emission of significant radiation. Radon contamination in the experimental environment presents a substantial background source in underground experiments. Therefore, it is crucial to measure the radon concentration.

%Uranium and thorium, among the materials that make up the Earth's surface, have radium among daughter nuclides. This radium decays to produce radon. Radon is a colorless, odorless, and tasteless radioactive noble gas. Radon repeatedly decays into lead, the most stable state, which produces a lot of radiation in the process. If the detector and the experiment environment are contaminated with radon, this radiation leaves a severe background on the detector. In addition, internal exposure to alpha decay through respiration or ingestion harms the organism's health. Therefore, it is essential to measure the radon concentration to understand the detector's background and construct a healthy experimental environment.

%Super-KamiokaNDE (Kamioka Neutrino Detection Experiment), Japan's representative neutrino experiment, is an underground experiment that observes neutrinos coming from the cosmos. About $\rm 50,000\ tons$ of ultra-pure water is being used as a target. In this experiment, a radon detector has been developed and operated since the 1990s to measure the level of contamination caused by radon \cite{Mitsuda2003}. 

Radon emanates from rocks, and as underground experimental facilities are encased within rock formations, there is a potential for radon contamination within the underground experimental facility. In such facilities, radon is removed by radon reduction systems (RRS), which reduce the radon concentration in the air by more than a hundred folds. Commercially available radon detectors cannot measure this low radon concentration. Therefore, a highly sensitive PIN photodiode-based radon detector has been designed and built for use at Yemilab.

\begin{figure}
    \centering
        \includegraphics{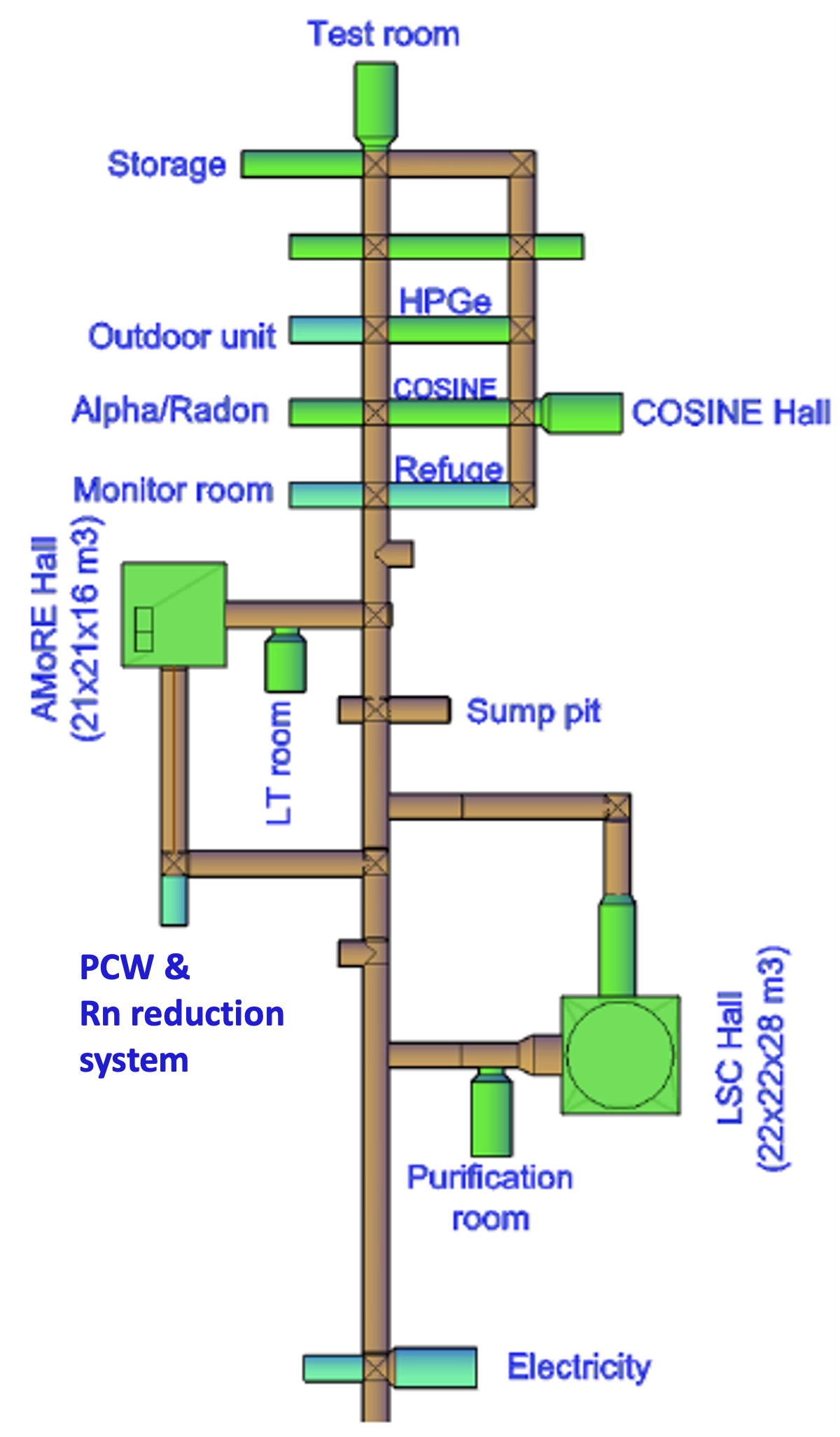}
    \caption{A schematic of the Yemilab experimental areas \cite{Park2021}.}
    \label{fig:SchemYemilab}
\end{figure}

%Yemilab introduction
Yemilab \cite{Park2021, Park2024, Lee2020_Yemi} is the new underground laboratory in Korea that has been operating since the end of 2022, which has 16 research spaces with a total area of $\rm 3000\ m^{2}$ and a 1000-meter overburden. The schematic of Yemilab is shown in Fig. \ref{fig:SchemYemilab}.
%Yemilab rock sample mention and cite
Yemilab has a cosmic muon background that is four times lower than the previous underground laboratory, Yangyang Underground Laboratory (Y2L) in Yangyang, Korea~\cite{Park2024}.

%Yemilab hosts several underground experiments in Korea.
%Several international and domestic experiments are being conducted in Yemilab.
The AMoRE \cite{AMoRETDR,  Lee2020} and COSINE \cite{Adhikari2018} experiments, two major physics experiments searching for neutrinoless double $\beta$ decay and weakly interacting massive particles, are being constructed.
Both experiments require extremely low background levels to observe very rare processes.
%\textcolor{red}{exceedingly rare processes. (exceedingly means "to a very great degree" jslee)}
It requires efforts to reduce the detector's intrinsic background and understand and reduce the background from the environment \cite{CHa2022}.
%The YeMiGO (Yemi Micro-Gravity Observatory) experiment \cite{Oh2023}, which measures microgravity fields, is conducted by the National Institute for Mathematical Sciences (NIMS). In addition, the Korea Atomic Energy Research Institute (KAERI) for precision $\gamma$ measurements, the Korea Institute of Geoscience and Mineral Resources (KIGAM) and Korea Meteorological Administration (KMA) for geological research, and Space LiinTech for space medicine are located in Yemilab.

%about RRS
Noble gases like radon bind weakly to materials. 
%due to the van der Waals force. 
To effectively adsorb radon in the air, a material with a large surface area is advantageous, and charcoal plays this role. A radon reduction system (RRS) is a device designed to reduce radon in the air by drawing radon-rich air and adsorbing the radon into the charcoal tank. As the charcoal tank's temperature decreases, the charcoal's capacity to adsorb the radon increases. Several RRS were developed in several underground laboratories and are still in operation today \cite{universe8020112, hodak2019characterization, murra2022design}.

The radon detector is designed to measure the concentration of radon that decays to airborne polonium in trapped air in a sampling chamber. 
Polonium becomes ionized and acquires a positive charge due to the recoil energy of its previous decay \cite{Agency1967}. Ionized polonium drifts into the photodiode following the electric field created by the voltage applied.
To achieve this, a sufficient volume is needed to contain sampled air and form an electric field to collect polonium in that air. From this perspective, a radon detector with a large air sample holding chamber was constructed,  similar to radon detectors currently being used in other underground laboratories \cite{Takeuchi1999, Wang1999, Hosokawa2015, Mitsuda2003, Flea2013, Gutirrez2004, Okamoto2021,
Elsio2020, Yakushev2010, Choi2001, Kiko2001, Mamedov2011}.
%https://doi.org/10.48550/arxiv.2112.06614, 

This paper discusses the radon concentration measurements in the air from the RRS with a highly sensitive radon detector at the Yemilab.

%Therefore, the shape of the radon chamber requires a large measurement volume capable of containing a lot of air. Also, the accuracy of the measurement data is proportional to this.
%Because the underground laboratory environment is surrounded by rocks, which are the source of radon gas, the radon gas is continuously supplied. This radon gas accumulates in the underground laboratory, distorting the experimental environment. A radon reduction system is needed to remove accumulated radon, which reduces the radon concentration in the air by factors between hundreds and thousands. This level cannot be measured with a typical radon meter. Therefore, a high-performance radon detector is required. Considering this situation, the measurement volume inside the radon detector was increased to enable the measurement of low radon concentrations and a silicon PIN photodiode was used as a sensor to ensure that the signal could be read.

%Radon detectors are closely related to the experimental point of view and the human living environment. In 2009, the World Health Organization (WHO) proposed a safety standard of 100 Bq/m3 for radon concentrations in residential areas \cite{WHObook}. In addition, as the danger of radon is highlighted and social interest increases, a commercial portable radon detector is being developed \cite{RAD7}.

%In this paper, we will describe the measurement results of a radon detector at the Yemilab.
%\clearpage
\section{Radon Reduction System}
\label{sec:rrs}

ATEKO \cite{webATEKO}, the most popular RRS manufacturer in the world, has provided the system at many underground facilities \cite{universe8020112}, including the one at Y2L. However, replacing the most high-pressure part of the system requires an additional certification by the Korea Occupational Safety and Health Agency (KOSHA). To avoid this vital regulation issue in Korea, we purchased an RRS manufactured by a Korean company specializing in ultra-pure water filtration systems.

\begin{figure}
    \centering
    \includegraphics[width=1\textwidth]{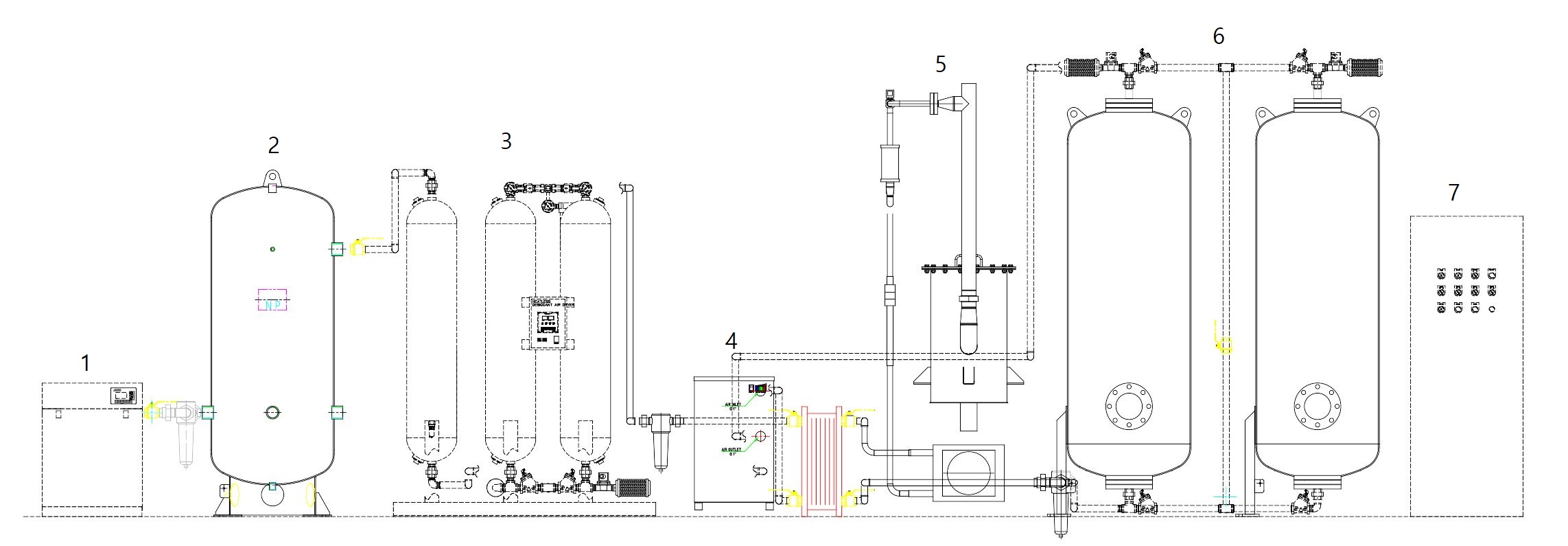}
    \caption{The layout of the custom-made Radon Reduction System. 1: Compressor, 2: Receiver tank, 3: Absorption dryer, 4: Air chiller, 5: Heater and HEPA filter units, 6: Active charcoal tanks, and 7: Control panel.}
    \label{fig:RRS}
\end{figure}

Figure \ref{fig:RRS} shows the layout of the RRS for the AMoRE-II at Yemilab. Air from the environment is supplied by the compressor and pressurized in the receiver tank. The absorption dryer can remove the moisture in the air under the $\rm -70^{\circ} C$ of dew point. This moisture-free pressurized air is contained in active charcoal tanks up to $\rm 10\ bars$. The temperature of active charcoal in the tanks has been cooled below $\rm -70^{\circ} C$ by the air chiller to remove radon from the dried air effectively. A heater and HEPA filter units are installed at the last part of the system to warm up and for dust filtration. The control panel operates all these processes automatically, controlled by a Windows-based program for online operation and monitoring.

The RRS aims to reduce the radon concentration below $\rm 1/1000$ that of the air in the Yemilab and supply the radon-free air with a capacity up to $\rm 60\ m^{3}/h$. A pre-tuning of the system was performed in the factory to optimize the operation conditions for a period longer than six months. It was decided to supply the air at $\rm 50\ m^{3}/h$ to minimize the maintenance effort.
%\clearpage
\section{Radon Detector}
\label{sec:analysis}

\begin{figure}
    \centering
    \includegraphics[width=0.3\textwidth]{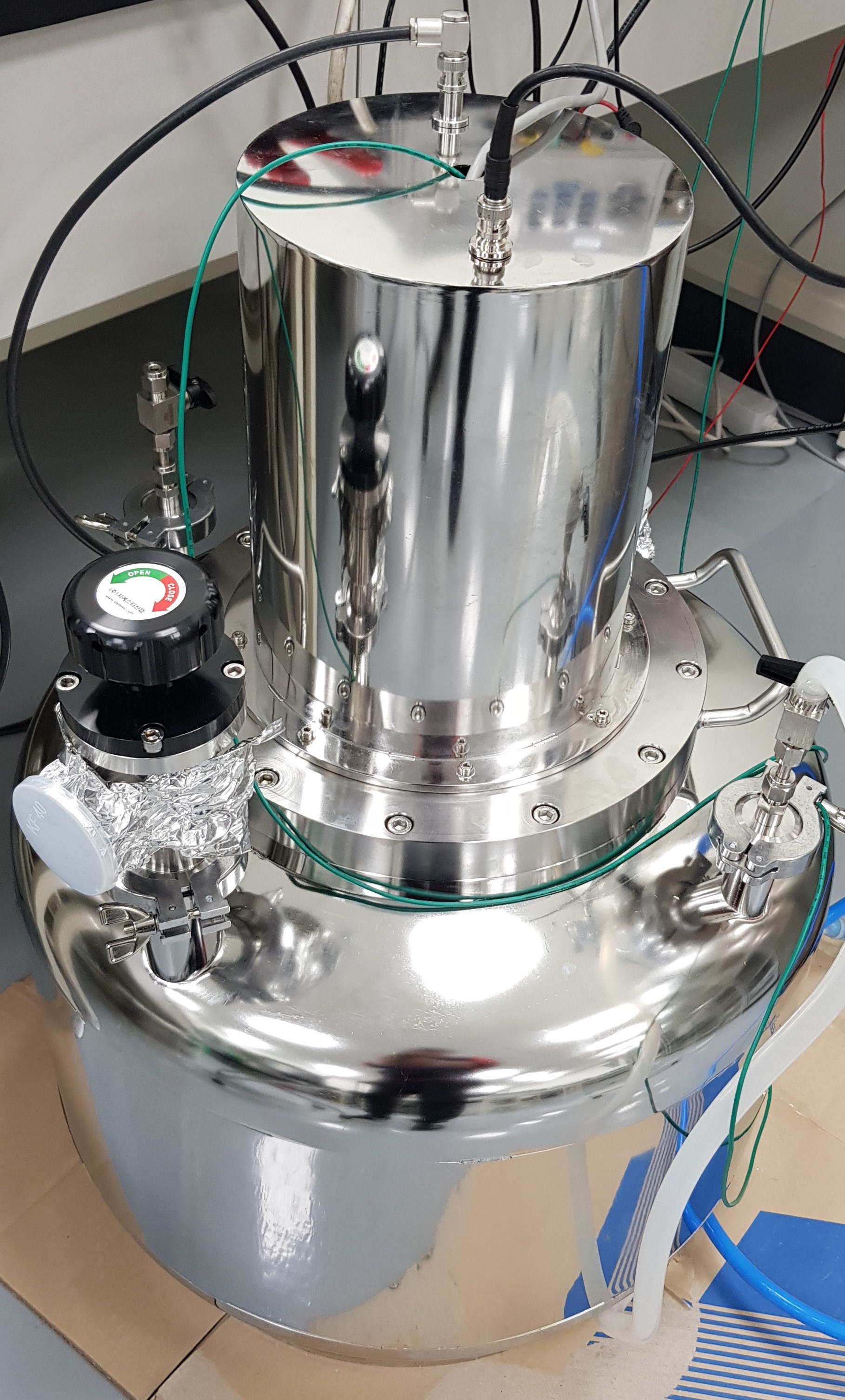}
    \hspace{15mm}
    \includegraphics[width=0.5\textwidth]{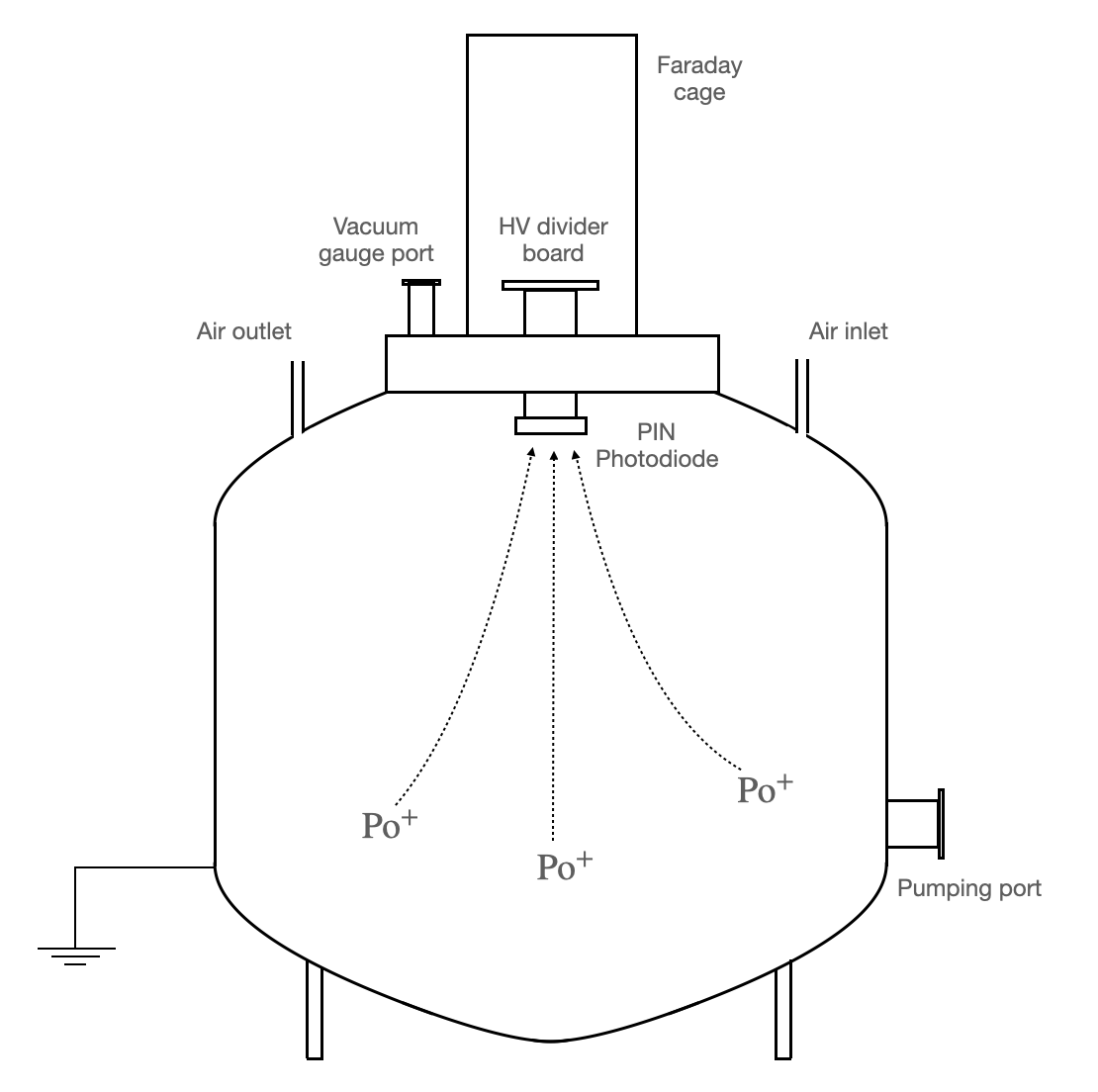}
    \caption{The picture (left) and schematic (right) of the radon detector. The measurement volume is the rounded part of the chamber below the flange plate. The measurement volume comprises $\rm 3\ mm$ thick stainless steel with an electropolished inner wall. The measurement volume is approximately $\rm 70\ L$. The detector has an air inlet, an air outlet, a port for the vacuum gauge, and a port connected to a vacuum pump to evacuate the internal air. There is a hole in the center of the flange for the feedthrough installation. Above the measurement volume is the front-end electronics board with voltage divider and preamplifier circuits. It is covered with a chamber lid. The chamber lid has three connector sockets, each with SHV, BNC, and power wires for the preamplifier. The outer wall of the chamber is connected to the ground.}
    \label{fig:RadonChamber}
\end{figure}

Figure~\ref{fig:RadonChamber} shows the radon detector developed for Yemilab.
The detector consists of a 70 L air sampling chamber with a PIN photodiode connected to a ceramic feedthrough mounted on the top flange. The chamber is made of 3~mm thick stainless steel, and the inner surface has been electrically polished to reduce the radon emanating from the chamber wall. The chamber has an inlet and an outlet to measure the air sample, and an additional vacuum pump connection is designed to evacuate the air inside the chamber quickly. The corners of the chamber are curved to form a uniform electric field, a different design from the radon chamber used in the radon monitoring for the KIMS experiment at Y2L~\cite{Lee2011, Lee2019}.

A windowless silicon PIN photodiode (S3204-09; Hamamatsu Photonics K.K.~\cite{s3204-09}) with dimensions of $18 ~\mathrm{mm} \times 18~ \mathrm{mm}$ is used.
A negative voltage is applied to the photodiode to establish an electric potential difference between the photodiode and the grounded inner surface of the chamber. Ionized polonium, a positively charged daughter particle from the radon decay, travels to the photodiode in the chamber.
%\textcolor{red}{Travel time of the ionized polonium is a few microsecond.}
Upon reaching the surface of the photodiode, polonium eventually decays and emits $\alpha$ particles. The emitted $\alpha$ particles lose energies in the photodiode's depletion region to create electron-hole pairs through which a signal is generated.

\begin{figure}
    \centering
    \includegraphics[width=1\textwidth]{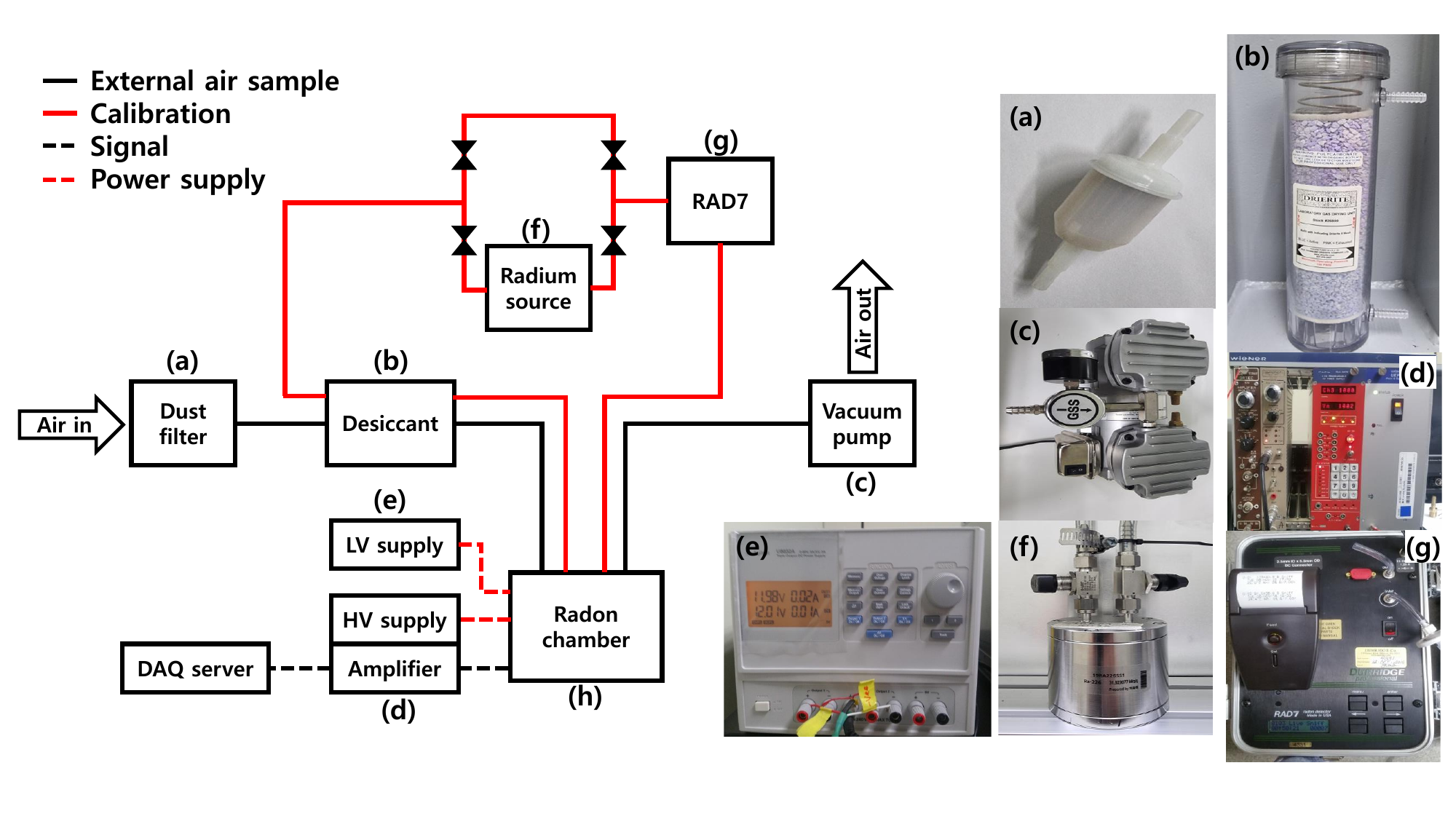}
    \caption{A schematic for the radon measurement setup is shown on the left. The solid (dashed) line is the flow line for the air (electrical connection). Other lines are explained in the legends on the top left. The pictures on the right show each part of the measurement setup, while (h) is the radon detector in Fig. \ref{fig:RadonChamber}.}
    \label{fig:Schem_measure}
\end{figure}

The air is introduced into the chamber via an oil-free vacuum pump (Rocker 400; Rocker Scientific Co. Ltd.~\cite{rocker400}), with a flowing rate of $\sim$ 20 L/min while keeping one atmospheric pressure. 
The dust in the sampled air is removed through a dust filter, which performs the same function as in the commercial radon detector (RAD7; Durridge Company, Inc.~\cite{rad7}). After filtration, the air undergoes dehumidification facilitated by a drying unit, a cylindrical acrylic container with approximately 500 g of desiccant (Drierite desiccant 23005~\cite{drierite}). The air passing through the desiccant has a relative humidity of 2 to 3$\%$RH.
These components are seamlessly connected using silicon tubes, as shown in Fig. ~\ref{fig:Schem_measure}.

\begin{figure}
    \centering
    \includegraphics*[width=1\textwidth, viewport=0 100 900 600]{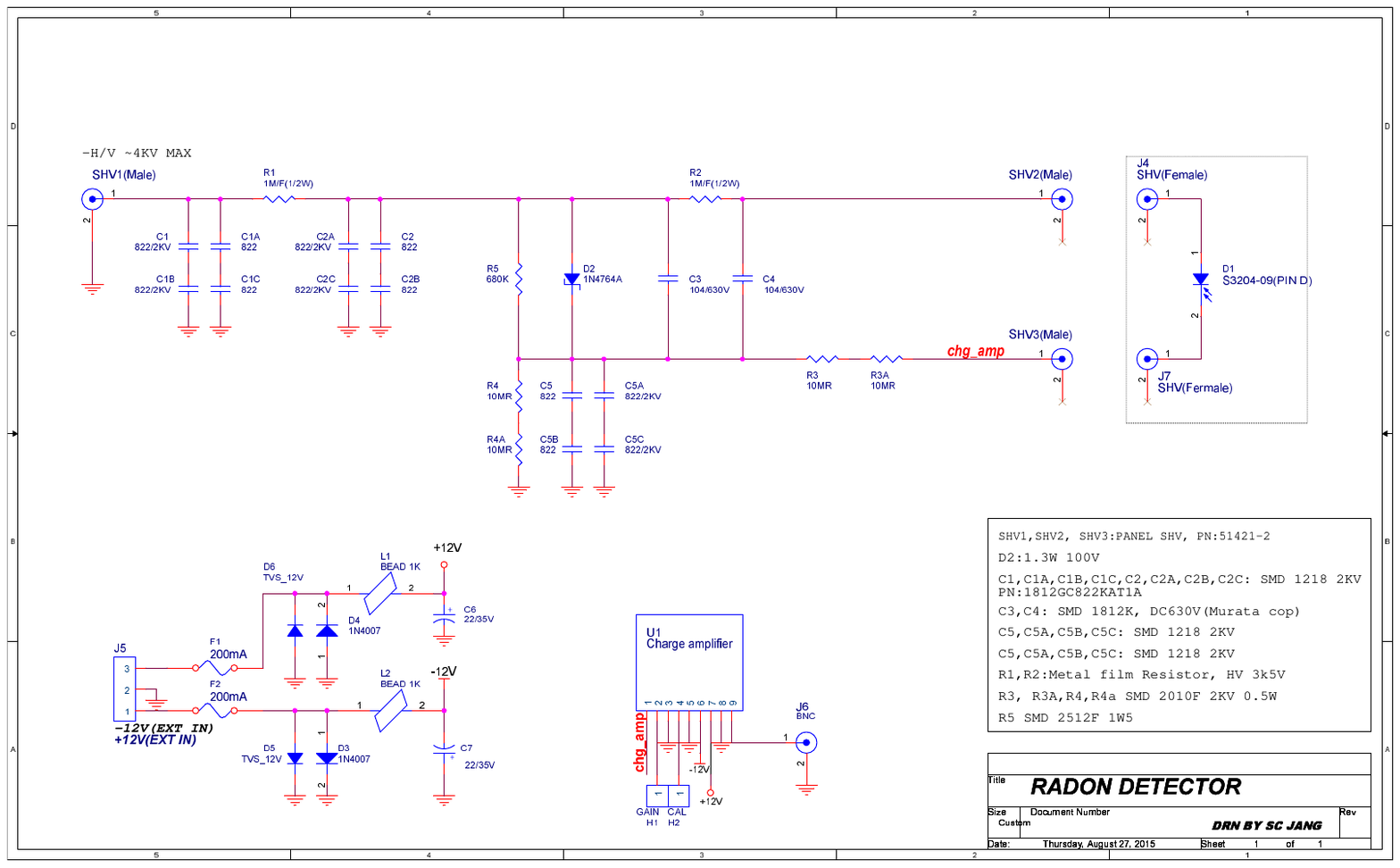}
    \caption{The schematic illustrates the front-end electronics, with the top circuit diagram depicting the HV divider section. This circuit is responsible for grounding the chamber and applying HV and bias to the photodiode. The schematics below detail the preamplifier circuit and its power supply. The photodiode signal goes into the preamplifier's first terminal for amplification, and the output is from terminal 9. This amplified signal is transmitted to the shaping amplifier through the BNC cable.}
    \label{fig:HVschem}
\end{figure}

The signal generated by the photodiode is sent to the front-end electronics board via a feedthrough. Figure~\ref{fig:HVschem} shows the circuit diagrams of the front-end electronics. The front-end electronics board has a preamplifier (H4083; Hamamatsu Photonics K.K.~\cite{h4083}) and a high-voltage (HV) divider to supply HV between the photodiode and the ground. There are BNC and SHV sockets for signal and HV, respectively. The front-end electronics board is covered with a Faraday cage to shield the external electromagnetic fields. 
On the left, a socket connects to the SHV cable, and on the right, there is a socket through which signals are transmitted through the BNC cable. The two holes in the center connect via a feedthrough to the photodiodes in the chamber. On the left side of the BNC connector, a board is designed to mount a preamplifier. This preamplifier is Hamamatsu's H4083 and is installed under a Faraday cage after soldering to the front-end electronics board to shield the external electromagnetic noise. The power to the preamplifier is supplied by the 3-pin connector at the bottom.
A HV supply (N470; CAEN Technologies, Inc.) is used to apply the HV for the radon detector and photodiode, and a low voltage (LV) supply (U8032A; Keysight Technologies, Inc.~\cite{u8032a}) to the preamplifier. 

\begin{figure}
    \centering
    \includegraphics[width=0.9\textwidth]{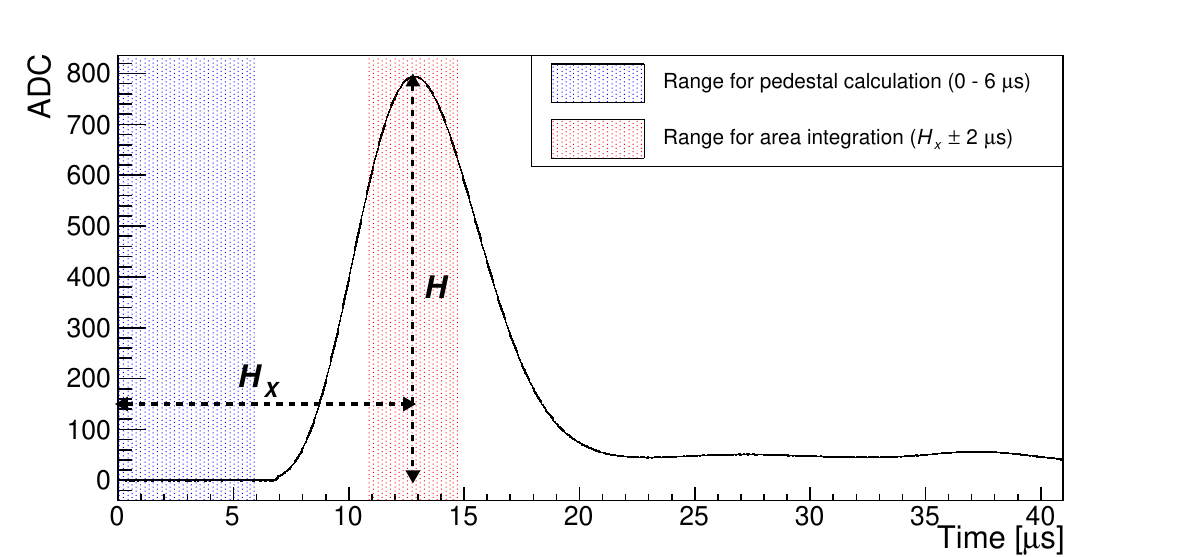}
    \caption{An example waveform of an $\alpha$ event after the pedestal subtraction.
    }
    \label{fig:waveform}
\end{figure}

Front-end electronics signals are shaped and amplified by a pulse-shaping amplifier (570; ORTEC Inc.~\cite{ortec570}). The shaping time used is $\rm 2\ \mu s$. The shaped signal is finally digitized by the 12-bit flash ADC with a sampling rate of 25 MS/s (NKFADC25; Notice Korea Co.). The pulse height trigger determines whether to save the waveform. Figure~\ref{fig:waveform} shows a typical waveform of an $\alpha$ signal.
All of the analysis parameters are reconstructed from the waveform.
Table~\ref{tab:parameters} defines six parameters used in the analysis.
\begin{table}[hbt!]
    \begin{center}
        \caption{The definition of parameters reconstructed from waveform.}
\vspace{3mm}
        \begin{tabular}{cc}
            \hline
            Parameter       & Description \\
            \hline
            $P\ {\rm and}\ P_{RMS}$ &  pedestal mean and RMS\\
            $H\ {\rm and}\ H_{x}$   &  maximum pulse height and its position.\\
            $A$                     &  integrated area ($H_{X}\pm 2~\rm{\mu} s$)\\
            $T_{\rm{trg}}$               &  trigger time\\
            \hline
            \label{tab:parameters}
        \end{tabular}
    \end{center}    
\end{table}

To correct the baseline of the waveform, the mean of the pedestal is calculated using the data in the 6 $\mu$s window before the pulse, and each ADC value of the waveform is subtracted by the mean of the pedestal, as shown in Fig.~\ref{fig:waveform}. The parameters corresponding to the energy of an event are $H$ and $A$. The $H$ is the maximum height of the waveform, and $A$ is the total ADC count calculated by integrating the waveform in the range of $H_{x}\pm 2 ~\mu$s. All of these have characteristics proportional to the signal's size (energy). The parameter $H$ is used as the corresponding energy value. The parameter $A$ is used primarily for noise rejection. The $P$ and $P_{\rm{RMS}}$ are calculated at the first $\rm 6\ \mu s$ time window of the waveform. In the case of short-term electrical noise or instability, these parameters tend to deviate from the nominal case. Some electrical noise signals are filtered through these parameters. 

\begin{figure}
    \centering
    \includegraphics[width=1\textwidth]{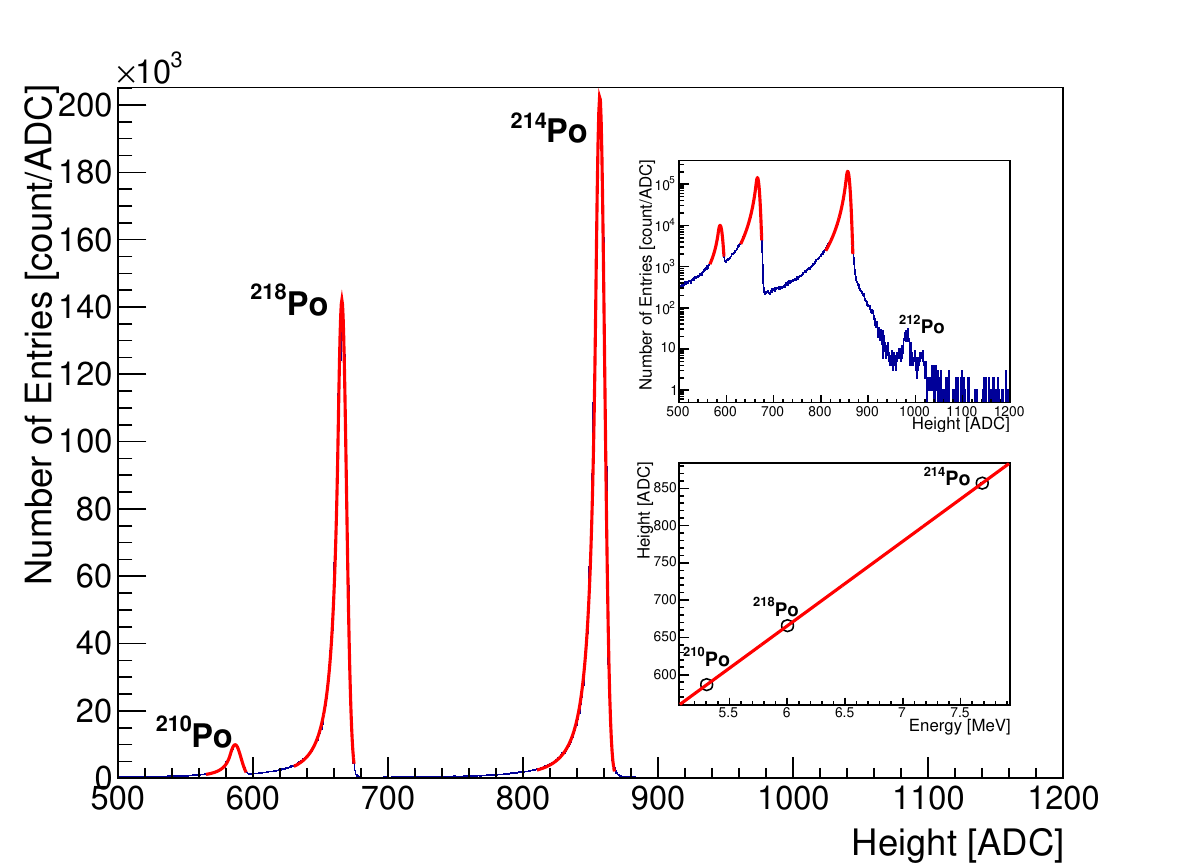}
    \caption{The ADC distribution with their fitting results. The ADC value is the difference between height and pedestal ($H-P$) for each event. Three peaks are shown in the figure in the order of $\rm ^{210}Po$, $\rm ^{218}Po$, and $\rm ^{214}Po$. Peaks in the $\rm ^{238}U$ chain are fitted with a Crystal Ball function, and the results are listed in Table \ref{tab:fitResult}.
    The upper part of the inset shows the logarithmic scale of the same distribution, indicating a relatively small $\rm ^{212}Po$ peak. The $\rm ^{212}Po$ has a short half-life of $\rm 299\ ns$ and most decay before reaching the diode, making the distribution different from other peaks. The energy calibration plot with a linear function fit of three peaks is below.}

    \label{fig:heightDist}
\end{figure}

As shown in Fig.~\ref{fig:heightDist}, four peaks are observed. The first three peaks ($\rm ^{212}Po,\ ^{218}Po,\ and\ ^{214}Po$) belong to the $\rm ^{238}U$ chain and the last peak is $\rm ^{212}Po$ from the $\rm ^{232}Th$ chain. Each peak, except for $\rm ^{212}Po$, is fitted with the Crystal Ball function [Eq.~\ref{eq:CrystallBall}], and the fitting results are summarized in Table~\ref{tab:fitResult}. 

\begin{equation}
\centering
    f(x; \alpha, n, \Bar{x}, \sigma)=\left\{
    \begin{matrix}
    \exp\left( - \frac{(x-\Bar{x})^{2}}{2 \sigma^{2}} \right),\ & {\rm for}\ \frac{x-\Bar{x}}{\sigma}>-\alpha \\
    A \cdot \left( B-\frac{(x-\Bar{x})}{\sigma} \right)^{-n},\ & {\rm for}\ \frac{x-\Bar{x}}{\sigma} \leq -\alpha
\end{matrix}
\right.
\label{eq:CrystallBall}
\end{equation}

The resolutions of all peaks ($\sigma/\Bar{x}$) are smaller than 1\%. $\rm ^{218}Po$ and $\rm ^{214}Po$, which are from the same decay chain of $\rm ^{222}Rn$, are used to estimate the concentration of radon. Due to its long half-life, 138 days, $\rm ^{210}Po$ is unsuitable for estimating the concentration.
$\rm ^{210}Pb$, which follows $\rm ^{214}Po$, has a half-life of about $\rm 22.3\ years$. Because of this characteristic, $\rm ^{210}Pb$ accumulates in the chamber and generates a signal. The rate of $\rm ^{210}Po$ decay from the $\rm ^{210}Pb$ is proportional to the age of the chamber. To eliminate this, the inside of the chamber must be electropolished, and the silicon diode must be replaced.
The $\rm ^{212}Po$ of the $\rm ^{232}Th$ decay chain has a different shape than the other polonium peaks. Due to the short half-life of $\rm 300\ ns$, $\rm ^{212}Po$ decays before reaching the photodiode and loses its recoil energy. 
Therefore, the locations of the three polonium peaks in the $\rm ^{238}U$ chain are used for the energy calibration through a linear function, as shown in Fig.~\ref{fig:heightDist}.

\begin{table}[]
    \centering
        \caption{The mean value ($\Bar{x}$) and $\sigma$ after Crystall Ball function fitting in Fig.~\ref{fig:heightDist}. The sigma resolution is calculated by $\sigma$ and $\Bar{x}$, and the FWHM is calculated by the histogram. $\rm{E}_\alpha$, alpha energy, values are from NNDC~\cite{nndc}.}
        \vspace{3mm}
    \begin{tabular}{c|c|c|c|c|c}
    \hline
    Isotopes & $\rm{E}_\alpha$ [MeV] & $\Bar{x}$ [ADC] & $\sigma$ [ADC]  & $\sigma/\Bar{x}$ [\%] & FWHM [keV]\\
    \hline \hline
    $\rm ^{210}Po$ & 5.304 & 587.506(33) & 4.214(37) & 0.72 & 81 \\
    $\rm ^{218}Po$ & 6.003 & 666.205(7)  & 3.275(5)  & 0.49 & 72 \\
    $\rm ^{214}Po$ & 7.687 & 857.450(5)  & 3.376(4)  & 0.39 & 72 \\
    \hline
    \end{tabular}
    \label{tab:fitResult}
\end{table}

\begin{figure}
    \centering
    \includegraphics[width=1\textwidth]{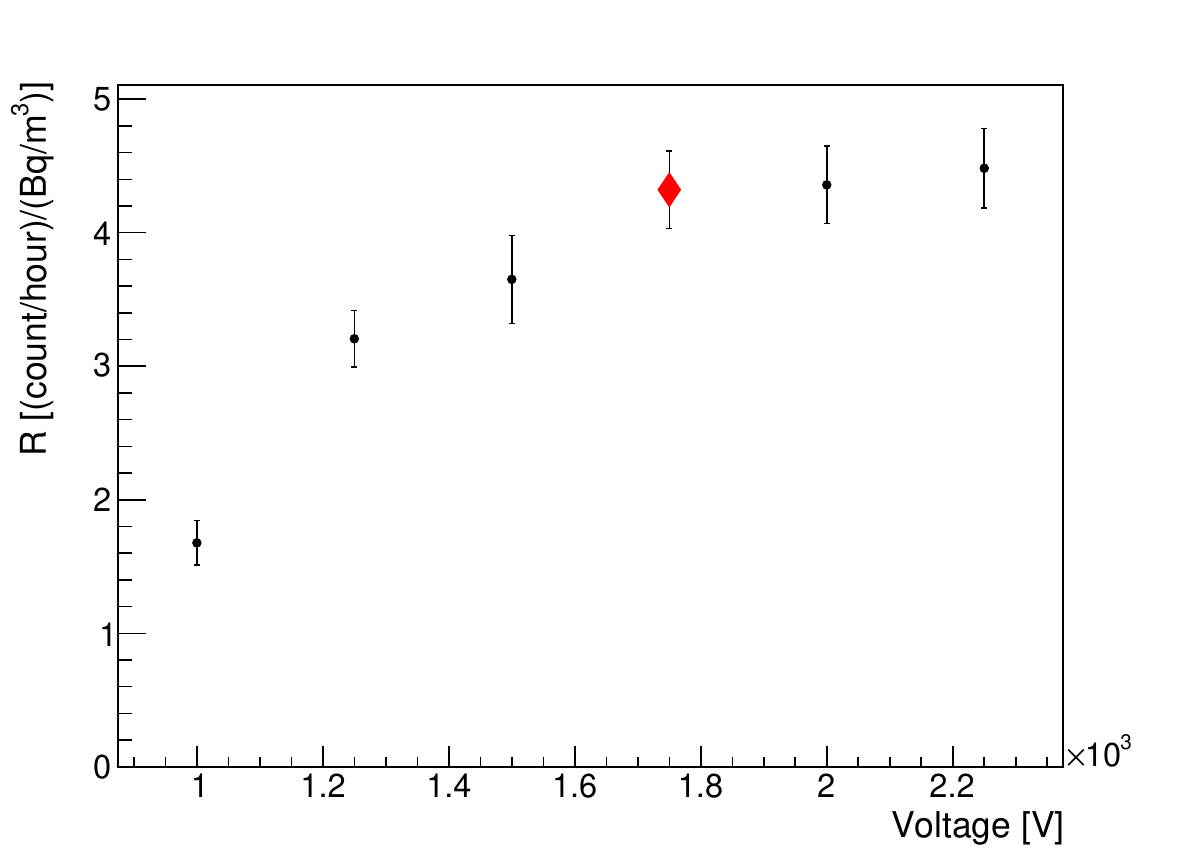}
    \caption{The activity ratios obtained from each data set from $1000~$ to $2250~\rm V$ with a step of $250~\rm V$. Each point is measured for about an hour. The location of the red diamond marker is where $R$ begins to saturate. This point is set as the operating voltage.}
    \label{fig:HVcal}
\end{figure}

To calibrate the HV and radon concentration of the radon detector, a closed loop system consisting of a radon detector, RAD7, and radium source is prepared as shown in Fig. \ref{fig:Schem_measure}. The efficiency of the radon detector depends on the HV applied. As the HV increases, the electric field for capturing polonium in the chamber intensifies. The efficiency is maximum when the voltage is high enough to capture all the polonium.
RAD7 has a hemispherical detection volume of $\rm 0.7\ L$ coated inside with an electrical conductor, and a silicon photodiode is installed at the center of the hemisphere.
The radium source, purchased from the Korea Research Institute of Standards and Science (KRISS), has an activity of approximately $\rm 1.5\ kBq$ and is sealed within a stainless steel container, as shown in Fig.~\ref{fig:Schem_measure}(f).
A filter is installed at the container outlet to prevent radium from being blown out directly.
Measurements are taken at six points by increasing $\rm 250\ V$ each time, from 1000 to $\rm 2250\ V$, and each measurement takes an hour.
The activity ratios ($R$) for HV are calculated, and its definition is
\begin{equation}
\centering
    R = A_{c}/A_{R},
\label{eq:CorrFact}
\end{equation}
where $A_c$ is the number of $^{218}$Po within $3 \sigma$ range of fitting result on an hour measurement, $A_R$ is the activity measured by the RAD7.
The $R$ versus HV is shown in Fig.~\ref{fig:HVcal}.
The operating voltage is determined at the point where $R$ starts to saturate, $\rm 1750\ V$.

Radon-rich air is injected into the closed-loop configuration to calibrate the detector's radon concentration.
In the closed loop of the radon detector and RAD7, as shown in Fig.~\ref{fig:Schem_measure}, the sampled air is circulated by the RAD7's pump (flowing rate of $\sim$ 0.8 L/min) with the source for approximately three hours to saturate it sufficiently.
After confirming that the radon concentration is saturated, the source is isolated from the loop. Radon remaining in the closed-loop structure naturally decays.
The radon detector and RAD7 measure the radon concentration in the closed loop simultaneously and continuously, which shows a reduction in radon levels. 
\begin{figure}
    \centering
    \includegraphics[width=1\textwidth]{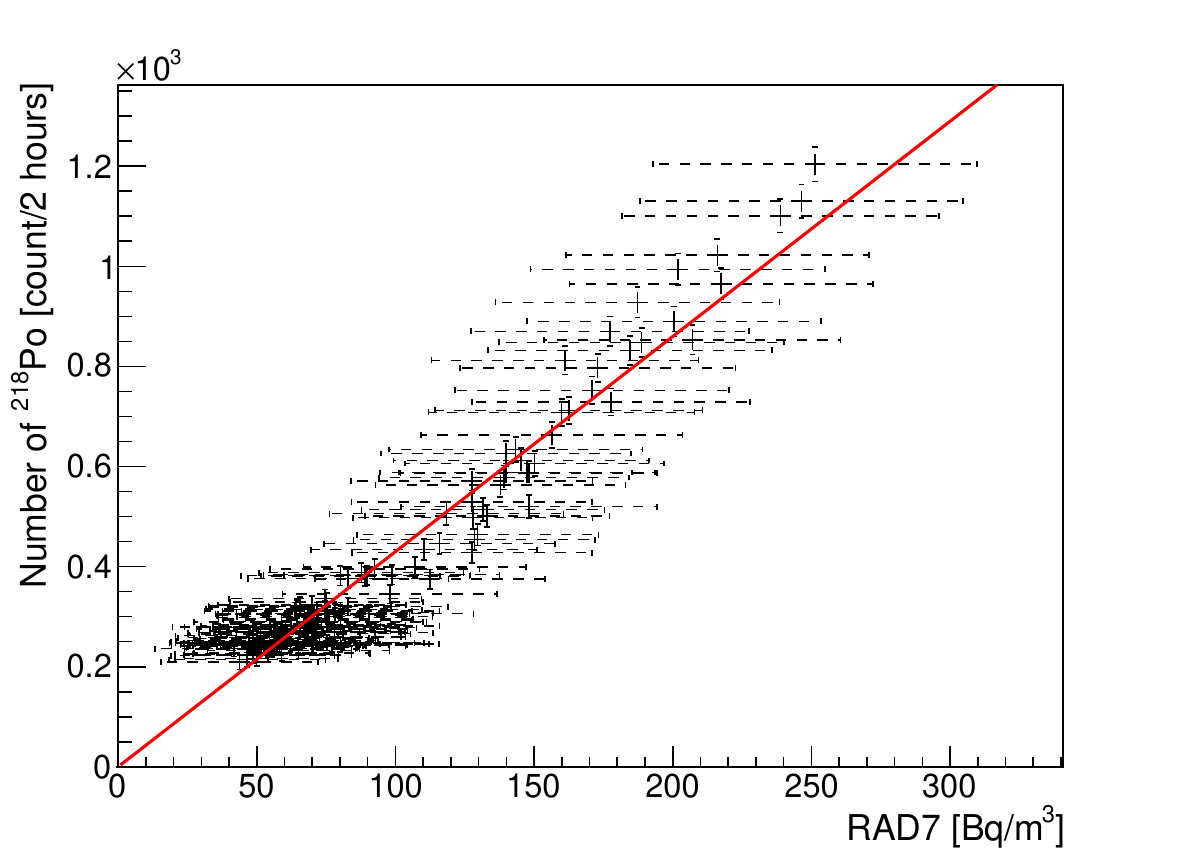}
    \caption{The correlation between RAD7 and radon detector after saturation of radon concentration. Over two weeks, the naturally declining radon concentration is measured using two detectors (RAD7 and the radon detector). A linear calibration function (red) without an offset fits the data.}
    \label{fig:ConCal}
\end{figure}
Figure~\ref{fig:ConCal} shows the correlation of radon concentration between RAD7 and the radon detector up to $\sim$~300 Bq/m$^3$ with about two weeks of data. Each point corresponds to data collected for 2 hours.
The correlation is fitted by a linear function without an offset, and the calibration function is obtained.

\begin{figure}
    \centering
    \includegraphics[width=1\textwidth]{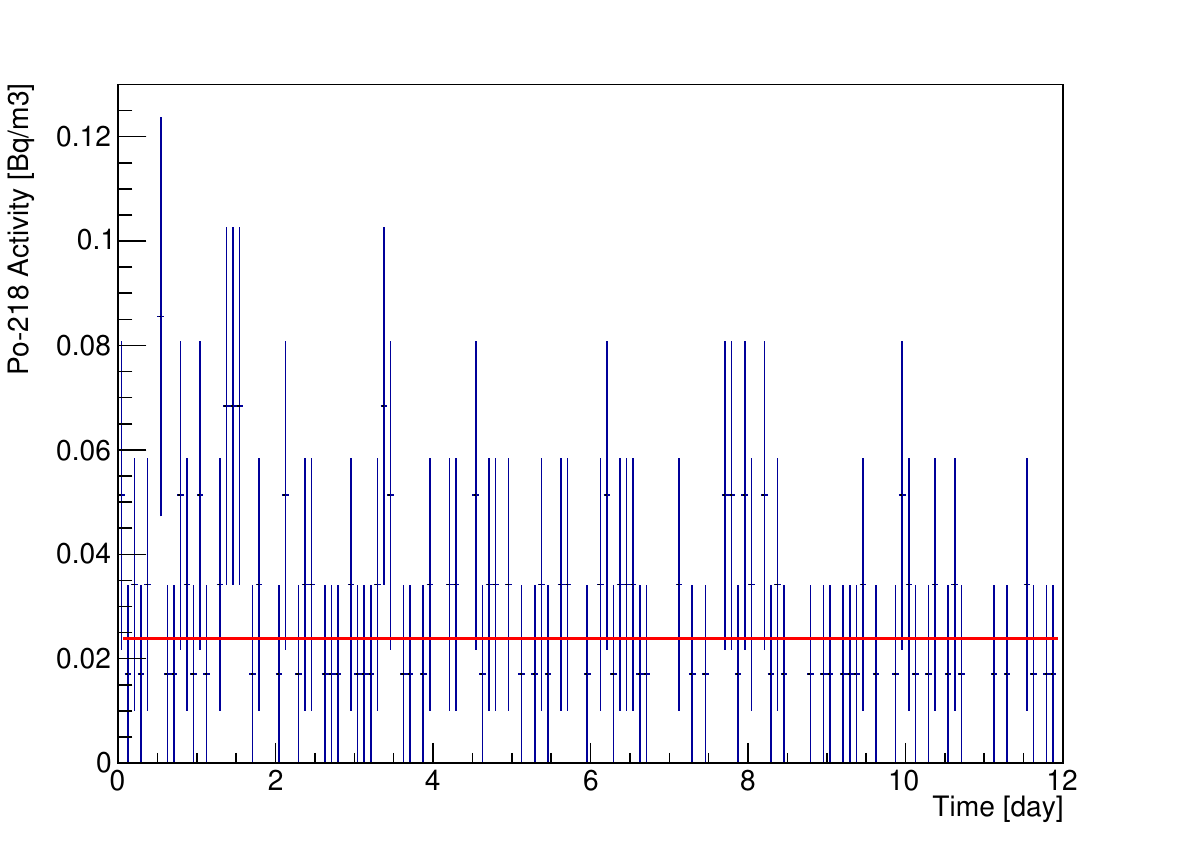}
    \caption{The intrinsic background measurement result of the radon detector. The radon detector was filled with pure boil-off nitrogen gas, and the average data value measured over 12 days is displayed as the red line, $\rm 23.8 mBq/m^3$}
    \label{fig:IntBKG}
\end{figure}

To understand the intrinsic background of the radon detector, about $\rm 15\ L/min$ of pure nitrogen gas from boil-off in a liquid nitrogen gas bottle is flushed to the chamber for $\rm 2\ hours$. The amount of nitrogen gas is enough to refresh the chamber volume approximately 25 times. The humidity within the chamber is about $\rm 40\ ppm$ when the chamber is sealed for measurement. Data are collected for approximately 12 days, and the intrinsic background of the radon detector is measured to be $\rm 23.8 \pm 2.1\ mBq/m^3$, as shown in Fig. \ref{fig:IntBKG}.

%\clearpage
\section{Radon Concentration Measurements}
\label{sec:measurement}
Environmental monitoring at Yemilab is in progress, including the concentration of radon.
%In the AMoRE-II experiment, an RRS was installed to reduce the radon concentration.
%This RRS is designed to supply $\rm 50\ m^{3}/h$ air at the $\rm 20\ mBq/kg$ level of radon \cite{yemi_inprep}. Two activated charcoal tanks operate using alternating cycles of adsorption and desorption. The cooled tank actively traps radon while the warmed tank releases accumulated gas, ensuring continuous removal. An air chiller was installed instead of a water chiller for the charcoal tank cooling for structural simplicity and system stability. 
%\textcolor{red}{Currently the RRS is operating $\rm 60\ m^{3}/h$ for load testing. (needed? jslee)}
Radon-reduced air by the RRS is piped to the AMoRE-II detector room and measured by the radon detector. Ambient air, before RRS, is measured by the RS9A, which is a compact pulsed-ionization type radon detector developed by RadonFTLab in South Korea \cite{webRS9A,Dimit}, and can measure the radon concentration below $\rm 3.7\ kBq/m^{3}$ with an accuracy of less than 15\%.
\begin{figure}
    \centering
    \includegraphics[width=1\textwidth]{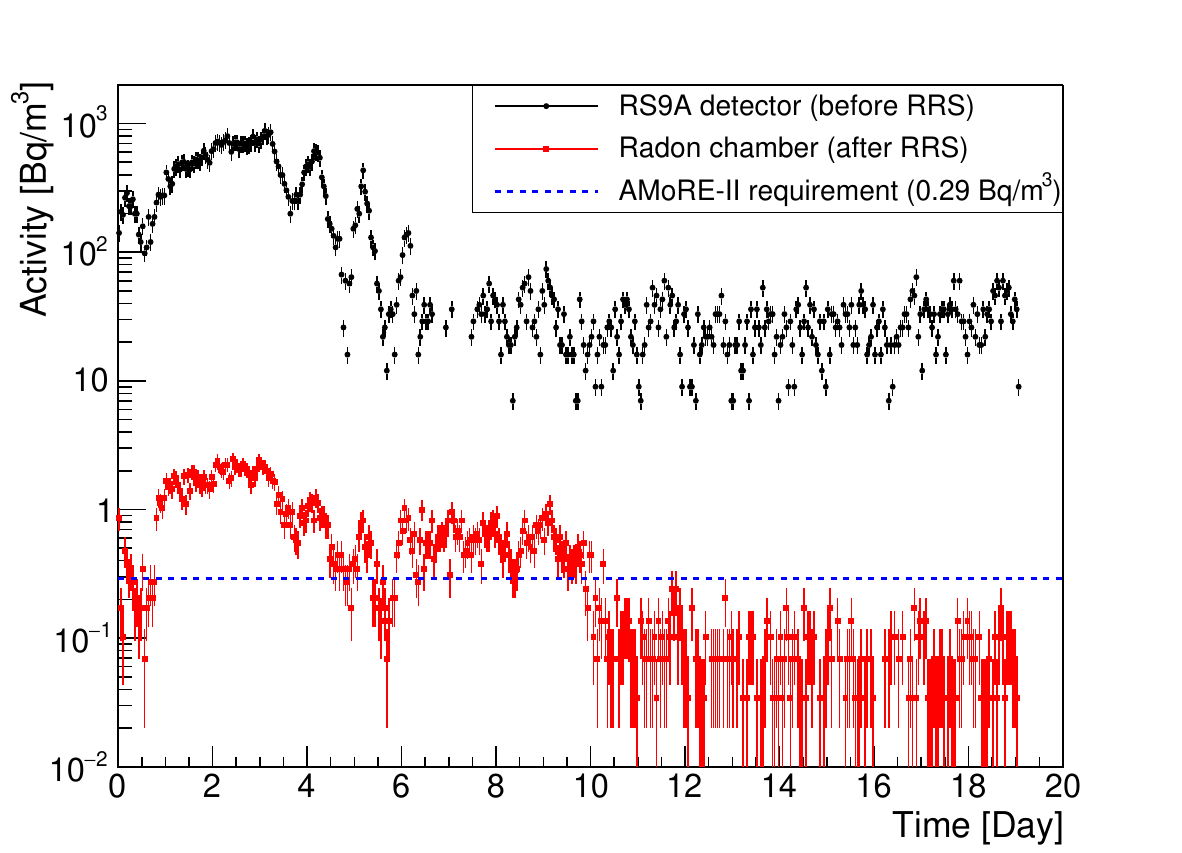}
    \caption{A comparison plot of radon concentration between the experiment hall air and the radon-reduced air from the RRS. Each data is measured by the radon detector (RS9A) for AMoRE's detector room air from the RRS (AMoRE experimental hall air to the RRS). It is assumed that the uncertainty of the RS9A data set is 15\% in precision. The blue dashed line is the AMoRE-II radon concentration requirement.}
    \label{fig:CompRRS}
\end{figure}
Figure \ref{fig:CompRRS} compares radon concentrations between the AMoRE experimental hall and the radon-reduced air from the RRS.
%During a total measurement period of about 64 days, an interruption occurred between the first and second data sets in the radon chamber for reasons such as RRS maintenance, power outage, and radon-reduced air supplies to another experimental space (i.e., the AMoRE-II clean room for crystal detector installation). In this period, the air displacement of RRS was reduced from $\rm 60 $ to $\rm 50\ m^{3}/h$.
The last nine days of data are used to estimate the reduction factor when the RRS is settled down.
The average radon concentration levels before and after RRS are $\rm 17.6 \pm 0.3$ and $\rm 0.058 \pm 0.004\ Bq/m^{3}$, respectively, and the reduction factor is about $1/300$.
The requirement of radon concentration for AMoRE-II is $\rm 0.29\ Bq/m^{3}$~\cite{amoreassay}, aligning with the region of interest for the neutrinoless double $\beta$ decay Q-value of $\rm ^{100}Mo$.
Figure \ref{fig:CompRRS} shows that the radon concentration satisfies the requirement after ten days.

The calibration process of the radon detector performed in this paper is a relative calibration using a RAD7 in a closed loop. Even though there is a desiccant in the loop, the RAD7 shows the humidity level to be 2 to 3$\%$RH, which is about $\rm 0.58\ g/m^{3}$. Since the humidity of the radon-reduced air from RRS is $\sim\rm 0.21\ g/m^{3}$, humidity is slightly different between the two setups. However, it seems negligible in calibration constant values between the two when looking at the humidity dependence of calibration constant in Fig. 5 of \cite{Hosokawa2015}. The RRS output measurement with lower humidity conditions could be overestimated due to a possible under-correction.

More rigorous calibration works with different humidity environments and flow rates will be carried out in the near future.

%\clearpage
\section{Summary}
\label{sec:conc}

%The radon detector collected data for about $\rm 180\ days$ in stable status. Bad events can be effectively removed with the waveform area parameter. Also, in the measured data, $\rm ^{210}Po$, $\rm ^{218}Po$, and $\rm ^{214}Po$ of the $\rm ^{238}U$ chain, and $\rm ^{212}Po$ of the $\rm ^{232}Th$ chain were observed. It was confirmed that each peak had an excellent resolution of less than 1$\%$ and good energy linearity.\\
The Yemilab, an underground laboratory in Jeongseon, Korea, was built for rare event search experiments to reduce cosmic-ray backgrounds. However, the Yemilab is still impacted by the radon background in the air. 

An RRS using a cold and pressurized charcoal chamber was designed and installed at the Yemilab to reduce the background radon in the AMoRE experiment, which is looking for the neutrinoless double beta decay from $^{100}$Mo isotopes in lithium molybdate crystal detectors at cryogenic temperatures. It reduces the radon level in the air at the experimental area to the level where the radon background becomes negligible to the experiment's sensitivity with a capacity of $\rm 50\ m^{3}/h$.

A highly sensitive radon detector with a 70 L chamber volume and a large silicon PIN photo-diode sensor readout was designed and manufactured to quantify the amount of radon reduced by the RRS. A calibration function for different HV values up to 2250 V was obtained using a closed loop connected with a calibrated commercial radon detector RAD7 using a radium source. Also, a radon activity calibration using the same setup was made up to $\sim$ 300 Bq/m$^3$ at 1750 V. From the calibration, the radon detector can measure alpha particles from $\rm ^{218}Po$ (6.003 MeV) and $\rm ^{214}Po$ (7.687 MeV), daughter isotopes of $\rm ^{222}Rn$, with an FWHM resolution of 72 keV. A boil-off N$_2$ gas test showed $\rm 23.8 \pm 2.1\ mBq/m^3$, the lowest background level indicating an intrinsic background. 

To confirm the reduction in radon by RRS, air from the RRS was measured using the radon detector, while the air going into the RRS was measured by another commercial radon detector, RS9A, for 19 days. From the measurement, a radon reduction factor of $\sim$ 300 by the RRS was measured. Also, the RRS output level reached below 0.29 Bq/m$^3$, which is the radon concentration requirement of AMoRE-II.

%The calibration process of the radon detector performed in this paper was relative calibration using RAD7. In addition, the humidity of the air in which radon was filtered through the RRS was $\rm 0.17\ g/m^{3}$. Still, the calibration was carried out in an environment with about 1 to 2$\%$$ relative humidity. This can be transferred to approximately $\rm 0.4\ g/m^{3}$. The difference in the correction factors can not be ignored \cite{Hosokawa2015}. More rigorous calibration work is required.
%Therefore, we will perform absolute calibration for humidity, flow rate, and radon concentration.

%Highlights
%\begin{itemize}
%    \item A radon detector has been constructed to monitor the radon concentration in the Yemilab environment.
%    \item Developed radon detector has a $\sigma$ resolution of less than 0.5\% for $\rm ^{218}Po$ and $\rm ^{214}Po$ peak, which is mainly used to estimate radon concentration, and the detection limit is XXX.
%    \item A radon reduction system (RRS) to remove the radon from the underground environment has been installed at the Yemilab, and the performance of the RRS has been confirmed to achieve the AMoRE-II requirement.
%\end{itemize}

%\clearpage
\acknowledgments
This work was supported by the Institute for Basic Science (IBS) funded by the Ministry of Science and ICT, Korea (Grant: IBS-R016-D1).
%\clearpage

%\acknowledgments
%This is the most common positions for acknowledgments. A macro is available to maintain the same layout and spelling of the heading.
%\paragraph{Note added.} This is also a good position for notes added after the paper has been written.

% Bibliography
%\bibliographystyle{JHEP}
%\bibliography{references}

\begin{thebibliography}{99}



\bibitem{Park2021}
K. S. Park et al., \emph{The new underground facility in Korea, Yemilab},
 \emph{Journal of Physics: Conference Series} {\bf 2156} 012171 (2021). 

\bibitem{Park2024}
K. S. Park et al., \emph{Construction of Yemilab},
 \emph{Front. Phys.} {\bf 12} 1323991 (2024). 
 %doi: 10.3389/fphy.2024.1323991

\bibitem{Lee2020_Yemi}
M. H. Lee \emph{Radioassay and Purification for Experiments at Y2L and Yemilab in Korea},
 \emph{Journal of Physics: Conference Series} {\bf 1468} 012249 (2020). 

\bibitem{AMoRETDR}
V. Alenkov et al. (AMoRE collaboration), 
\emph{Technical Design Report for the AMoRE $0\nu\beta\beta$ Decay Search Experiment},
arxiv:1512.05957 (2015).
%https://doi.org/10.48550/arxiv.1512.05957

\bibitem{Lee2020}
M. H. Lee (AMoRE collaboration), \emph{AMoRE: a search for neutrinoless double-beta decay of $^{\rm 100}$Mo using low-temperature molybdenum-containing crystal detectors},
 \emph{JINST} {\bf 15(08)} C08010 (2020). 

\bibitem{Adhikari2018}
G. Adhikari et al. (COSINE collaboration), \emph{Initial performance of the COSINE-100 experiment},
 \emph{The European Physical Journal C} {\bf 78} 107 (2018). 
%  url = {http://dx.doi.org/10.1140/epjc/s10052-018-5590-x},
%  DOI = {10.1140/epjc/s10052-018-5590-x},

\bibitem{CHa2022}
C. Ha et al., \emph{Radon concentration variations at the Yangyang underground laboratory},
 \emph{Front. Phys.} {\bf 10} 1030024 (2022). 
 %https://doi.org/10.3389/fphy.2022.1030024

\bibitem{universe8020112}
J. Pérez-Pérez et al., \emph{Radon Mitigation Applications at the Laboratorio Subterráneo de Canfranc (LSC)},
 \emph{Universe} {\bf 8(2)} 112 (2022). 
%URL = {https://www.mdpi.com/2218-1997/8/2/112},
%DOI = {10.3390/universe8020112}
 
\bibitem{hodak2019characterization}
R. Hod{\'a}k et al., \emph{Characterization and long-term performance of the Radon Trapping Facility operating at the Modane Underground Laboratory},
 \emph{Journal of Physics G: Nuclear and Particle Physics} {\bf 46(11)} 115105 (2019). 
%Citation R Hodák et al 2019 J. Phys. G: Nucl. Part. Phys. 46 115105
%DOI 10.1088/1361-6471/ab368e

\bibitem{murra2022design}
M. Murra et al., \emph{Design, construction and commissioning of a high-flow radon removal system for XENONnT},
 \emph{The European Physical Journal C} {\bf 82(12)} 1104 (2022). 
%Murra, M., Schulte, D., Huhmann, C. et al. Design, construction and commissioning of a high-flow radon removal system for XENONnT. Eur. Phys. J. C 82, 1104 (2022). 
%https://doi.org/10.1140/epjc/s10052-022-11001-9

\bibitem{Agency1967}
E. Albrecht and A. Kaul, \emph{Continuous registration of $^{222}$Rn concentration in air varying with time (SM-95/39)}, \emph{Assessment of Airborne Radioactivity, Proceeding of IAEA Symposium (Vienna, 3-7 July, 1967) 643 (1967)}.

\bibitem{Takeuchi1999}
Y. Takeuchi et al., \emph{Development of high sensitivity radon detectors},
 \emph{Nuclear Instruments and Methods in Physics Research Section A: Accelerators,  Spectrometers,  Detectors and Associated Equipment} {\bf 421} 334 (1999). 
%https://doi.org/10.1016/S0168-9002(98)01204-2

\bibitem{Wang1999}
J.-X. Wang et al., \emph{An electrostatic radon detector designed for water radioactivity measurements},
 \emph{Nuclear Instruments and Methods in Physics Research Section A: Accelerators,  Spectrometers,  Detectors and Associated Equipment} {\bf 421} 601 (1999). 
%https://doi.org/10.1016/S0168-9002(98)01230-3

\bibitem{Hosokawa2015}
K. Hosokawa et al., \emph{Development of a high-sensitivity 80 L radon detector for purified gases},
 \emph{Progress of Theoretical and Experimental Physics} {\bf 2015} 033H01 (2015). 
%Progress of Theoretical and Experimental Physics, Volume 2015, Issue 3, March 2015, 033H01, 
%https://doi.org/10.1093/ptep/ptv018

\bibitem{Mitsuda2003}
C. Mitsuda et al., \emph{Development of super-high sensitivity radon detector for the Super-Kamiokande detector},
 \emph{Nuclear Instruments and Methods in Physics Research Section A: Accelerators,  Spectrometers,  Detectors and Associated Equipment} {\bf 497} 414 (2003). 
%https://doi.org/10.1016/S0168-9002(02)01923-X

\bibitem{Flea2013}
S. Folea et al., \emph{WI-FI portable solution for distributed radon measurements},
 \emph{Romanian Journal of Physics} {\bf 58} S126 (2013). 
%https://rjp.nipne.ro/2013_58_Suppl/0126_0139.pdf

\bibitem{Gutirrez2004}
J.L. Gutiérrez et al., \emph{Radon emanation measurements using silicon photodiode detectors},
 \emph{Applied Radiation and Isotopes} {\bf 60} 583 (2004). 
%https://doi.org/10.1016/j.apradiso.2003.11.080

\bibitem{Okamoto2021}
K. Okamoto et al., 
\emph{Improvement of radon detector performance by using a large-sized PIN-photodiode},
arxiv:2112.06614 (2021).
%https://doi.org/10.48550/arxiv.2112.06614,
%doi = {10.48550/ARXIV.2112.06614},

\bibitem{Elsio2020}
S. Elisio and L. Peralta, \emph{Development of a low-cost monitor for radon detection in air},
 \emph{Nuclear Instruments and Methods in Physics Research Section A: Accelerators,  Spectrometers,  Detectors and Associated Equipment} {\bf 969} 164033 (2020). 
 %url = {http://dx.doi.org/10.1016/j.nima.2020.164033},
 %DOI = {10.1016/j.nima.2020.164033},

\bibitem{Yakushev2010}
E.A. Yakushev and A.V. Lubashevskii, \emph{Radon-induced background and methods of its elimination in the EDELWEISS-II experiment},
 \emph{Russian Physics Journal} {\bf 53} 616 (2010). 
%  url = {http://dx.doi.org/10.1007/s11182-010-9462-6},
%  DOI = {10.1007/s11182-010-9462-6},

\bibitem{Choi2001}
E. Choi et al., \emph{Highly sensitive radon monitor and radon emanation rates for detector components},
 \emph{Nuclear Instruments and Methods in Physics Research Section A: Accelerators,  Spectrometers,  Detectors and Associated Equipment} {\bf 459} 177 (2001). 
%  url = {http://dx.doi.org/10.1016/S0168-9002(00)01004-4},
%  DOI = {10.1016/s0168-9002(00)01004-4},

\bibitem{Kiko2001}
J. Kiko, \emph{Detector for $^{222}Rn$ measurements in air at the 1 mBq/m$^3$ level},
 \emph{Nuclear Instruments and Methods in Physics Research Section A: Accelerators,  Spectrometers,  Detectors and Associated Equipment} {\bf 460} 272 (2001). 
%  url = {http://dx.doi.org/10.1016/S0168-9002(00)01082-2},
%  DOI = {10.1016/s0168-9002(00)01082-2},


\bibitem{Mamedov2011}
F. Mamedov et al., \emph{Development of an ultra-sensitive radon detector for the SuperNEMO experiment},
 \emph{JINST} {\bf 6(12)} C12008 (2011). 
%  url = {http://dx.doi.org/10.1088/1748-0221/6/12/C12008},
%  DOI = {10.1088/1748-0221/6/12/c12008},

\bibitem{webATEKO}
ATEKO a.s., \emph{https://www.ateko.cz/en},
 (Accessed February 15, 2024).

\bibitem{Lee2011}
M. Lee et al., \emph{Radon Environment in the Korea Invisible Mass Search Experiment and Its Measurement},
 \emph{Journal of the Korean Physical Society} {\bf 58} 713 (2011). 
%  doi = {10.3938/jkps.58.713},
%  url = {https://doi.org/10.3938/jkps.58.713},

\bibitem{Lee2019}
K. M. Seo et al., \emph{Silicon PIN photodiode-based radon detectors for an underground experiment environment},
 \emph{Proceedings of The 39th International Conference on High Energy Physics — PoS(ICHEP2018)} {\bf 340} 803 (2019). 
%  url = {http://dx.doi.org/10.22323/1.340.0803},
%  DOI = {10.22323/1.340.0803},

\bibitem{s3204-09}
Hamamatsu silicon photodiode S3204-09, \emph{https://www.hamamatsu.com/us/en/product/optical-sensors/photodiodes/si-photodiodes/S3204-09.html},
 (Accessed February 16, 2024).

\bibitem{rocker400}
Rocker 400 vacuum pump, \emph{https://www.rocker.com.tw/en/product/lab-pumps/oil-free-vacuum-pump/rocker-400-oil-free-vacuum-pump/},
 (Accessed February 16, 2024).

\bibitem{rad7}
RAD7 RADON DETECTOR, \emph{https://durridge.com/products/rad7-radon-detector/},
 (Accessed February 16, 2024).
 
\bibitem{drierite}
DRIERITE 23005, \emph{https://secure.drierite.com/catalog3/product.cfm?item{\_}num=23005/},
 (Accessed February 16, 2024).

\bibitem{h4083}
Hamamatsu charge amplifier H4083, \emph{https://www.hamamatsu.com/content/dam/hamamatsu-photonics/sites/documents/99{\_}SALES{\_}LIBRARY/ssd/charge{\_}amp{\_}kacc9001e.pdf},
 (Accessed February 16, 2024).

\bibitem{u8032a}
Keysight Triple Output DC Power Supply U8032A, \emph{https://www.keysight.com/us/en/product/U8032A/triple-output-dc-power-supply-60v-3a-2x-5v-3a-375w.html},
 (Accessed February 16, 2024).

\bibitem{ortec570}
Ortec amplifier 570, \emph{https://www.ortec-online.com/-/media/ametekortec/brochures/5/570.pdf?la=en\&revision=53f9a77c-d8f8-4459-80b8-a93822b2aab2},
(Accessed February 16, 2024).

\bibitem{nndc}
National Nuclear Data Center, \emph{https://www.nndc.bnl.gov/},
 (Accessed February 15, 2024).

\bibitem{webRS9A}
IoT radon sensor: RS9A, \emph{http://radonftlab.com/kr/iot-radon-sensor-rs9a/},
 (Accessed February 15, 2024).

\bibitem{Dimit}
I. Dimitrova et al., \emph{Study of the performance and time response of the RadonEye Plus2 continuous radon monitor},
 \emph{Measurement} {\bf 207} 112409 (2023). 
%https://doi.org/10.1016/j.measurement.2022.112409

\bibitem{amoreassay}
A. Agrawal et al. (AMoRE collaboration), \emph{Radioassay of the materials for AMoRE-II experiment},
 \emph{Submitted to Front. Phys.} {\bf} (2024). 


%\bibitem{WHObook}
%WHO handbook on indoor radon: a public health perspective, \emph{https://www.who.int/publications/i/item/9789241547673},
% (Accessed February 15, 2024).

%\bibitem{Oh2023}
%J.J. Oh et al., \emph{New Deep Underground Microgravity Laboratory in South Korea},
% \emph{EGU General Assembly 2023, Vienna, Austria, 23–28 Apr 2023,} {\bf EGU23} 10359 (2023). 
% https://doi.org/10.5194/egusphere-egu23-10359, 2023.


%\bibitem{a}
%Author, \emph{Title}, \emph{J. Abbrev.} {\bf vol} (year) pg.

%\bibitem{b}
%Author, \emph{Title},
%arxiv:1234.5678.

%\bibitem{c}
%Author, \emph{Title},
%Publisher (year).

\end{thebibliography}

\end{document}